\documentclass[a4paper, 12pt]{article}

\usepackage{graphicx}
\usepackage{url}
\usepackage{amsbsy,amssymb}
\usepackage{amsmath}
\usepackage{abstract}
\usepackage{hyperref}
\usepackage{color}

\usepackage{amsmath,amsthm}
\usepackage{mathtools}
\usepackage{xfrac}
\usepackage{faktor}

\usepackage[all]{xy}

\usepackage{orcidlink}

\newtheorem{Proposition}{Proposition}
\newtheorem{Corollary}{Corollary}

\newtheorem{Theorem}{Theorem}
\newtheorem*{Proof}{Proof}

\newtheorem{Definition}{Definition}

\newtheorem{Example}{Example}
\newtheorem{Algorithm}{Algorithm}

\newcommand{\dcov}{d^{\nabla}}

\begin{document}

{\LARGE\centering{\bf{Covariant tomography of fields}}}

\begin{center}
\sf{Rados\l aw Antoni Kycia$^{1,a}$  \orcidlink{0000-0002-6390-4627}}
\end{center}

\medskip
 \small{
\centerline{$^{1}$Cracow University of Technology}
\centerline{Faculty of Computer Science and Mathematics}
\centerline{Warszawska 24, Krak\'ow, 31-155, Poland}
\centerline{\\}

\centerline{$^{a}${\tt
kycia.radoslaw@gmail.com} }
}

\begin{abstract}
\noindent
This paper develops 'covariant tomography', a local framework for solving Inverse Boundary Value Problems (IBVP) for parallel transport equation on star-shaped domains. By integrating geometric decomposition with specific interior extensions - radial, heat equation, or harmonic - the method reconstructs currents and gauge potentials from boundary data. The choice of extension directly dictates the regularity of the recovered interior fields. A primary contribution is the 'tower' algorithm, which reduces higher-order systems, such as Maxwell equations, to a sequence of coupled first-order equations. We establish a formal solvability criterion (Theorem \ref{Th_TowerTheorem}), proving that higher-order IBVPs are solvable if and only if this tower is sequentially solvable. The framework is validated through low-dimensional examples and electromagnetic potential reconstruction in $\mathbb{R}^{3}$.
\end{abstract}

\section{Introduction}
A typical boundary value problem (BVP) for differential equations can be formulated in an abstract way as follows \cite{GeneralizedInverseFredholm}. For a vector spaces $V$, $W$ and operators $L:V\rightarrow V$ and $l:V\rightarrow W$ we have to solve for $y$ a system
\begin{eqnarray}
    Ly = z \\
    ly = \lambda,
    \label{Eq.AbstractBVP}
\end{eqnarray}
where $y,z \in V$ and $\lambda \in W$. The operator $L$ is typically a differential operator and $l$ is a 'trace' operator that maps a solution $y$ to the boundary value $\lambda \in W$. 

Historically, the problem was formulated in \cite{Gelfand} for the inverse Sturm-Liouville problem and pursued in \cite{Novikov} for the multidimensional inverse problem for the Schr\"{o}dinger equation. The specific problems for ODEs, finite difference, and integral equations \cite{GeneralizedInverseFredholm}, PDEs \cite{Showalter}, \cite{Grubb_Operators}, or, in particular, exterior differential equations \cite{Schwarz_exteriorDerivative} can be defined.

The algorithm for solving (\ref{Eq.AbstractBVP}) consists of two general steps:
\begin{enumerate}
    \item {Solve the BVP for $\lambda=0$ constructing a Green function yielding general solution typically parametrized.}
    \item {Use a 'pseudoinverse' of $l$ (if it exists) to fix the general solution/Green's function to restore the specific boundary condition $y$.}
\end{enumerate}
The pseudoinverse arises because $l$ is non-bijective and typically 'singular'; e.g., in the setting of Fredholm operators in Banach/Hilbert spaces, the kernel of $l$ can be large. Therefore, this is a major obstacle to effectively solving BVP. Therefore, to fix this, the operator $L$ itself (its internal structure) must be varied. This influences the non-uniqueness of restoring the form of the operator in the interior of a region, knowing only the solution at the boundary - a tomography or Inverse Boundary Value Problem (IBVP), see \cite{InverseProblemsInDE}. This is a problem related to the famous Mark Kac question, 'Can one hear the shape of a drum?' (isospectral problem) \cite{Kac_drum} that was solved in \cite{drum_final_solution}, see also \cite{Milnor_drum}.

The IVBP can be characterized by the following: 'the problem of determining unmeasurable parameters of a system from measurable parameters, by using a mathematical/physical model' \cite{InverseProblemsInDE}. Since the operator shape encodes the physics of the problem, in other words, we try to recover the physical properties of the problem from measured 'boundary' values.

One of the important IBVPs in physics is to determine the conditions (gauge fields, currents) such that, when plugged into the parallel transport (covariant constancy) equation within a specific region, they yield a specific value at the boundary. The solution of such problems, as in medical tomography, allows us to infer the properties of the region's interior from knowledge only of the solution outside. As discussed above, we cannot, in general, expect to recover an exact solution and the region parameters (current, gauge potential) inside, and only some solution classes are given, i.e., we expect non-uniqueness in tomography.

The problem of parallel transport can be formulated in terms of exterior calculus, which allows one to use various powerful techniques and algorithms \cite{Bleecker, Nakahara, Thirring, Frankel}. In this paper, we used one such method, 'geometric decomposition' \cite{EdelenExteriorCalculus, EdelenIsovectorMethods}, which employs the Poincar\'{e} lemma and homotopy operator properties to treat exterior differential equations (PDEs) in analogy to ODEs \cite{KyciaSilhan}. The only drawback is that the underlying homotopy operator requires a simple topology of the underlying domain - a star-shaped/contractible open subset. Yet, for various practical applications it is sufficient \cite{KyciaSilhan, KyciaPoincare, KyciaPoincareCohomotopy, CopoincareHamiltonianSystem}.

Geometric methods in IBVP play increasing role and define various directions which we summarize below:
\begin{itemize}
    \item {\textbf{Calder\'{o}n problem} related to Electrical Impedance Tomography was initiated in \cite{CalderonProblem0, CalderonProblem2}, see also the review in \cite{CalderonProblem1} and applications to recovery of Riemannian manifold in \cite{CalderonProblem3}.}
    \item {\textbf{Boundary Control Method (BCM)} was formulated in \cite{BCM0}, see also the review in \cite{BCM1}, and applications to Riemannian manifolds in \cite{BCM2, BCM3, BCM4}.}
    \item {\textbf{Geodesic Ray Transform} proposed in \cite{GRT0}, for further developement see \cite{GRT1, GRT2, GRT3}.}
    \item {\textbf{Boundary Rigidity Problem} formulated in \cite{BRP0}, see further developed in \cite{BRP1, BRP2}.}
    \item {Electromagnetic problems can be treated by Inverse Scattering Problem \cite{Cakoni_EM, Cakoni_ISP}.}
    \item {\textbf{Inverse Problems in General Relativity} proposed in \cite{IGR0} and furthr developed in \cite{IGR1, IGR2}.}
\end{itemize}
We propose a new approach that can be used to IVBP that are formulated in terms of PDEs on manifolds employing the properties of differential operators.

The paper is organized as follows: In the next section, we summarize all necessary facts about geometric decomposition and the parallel transport equation. Then we describe our results on covariant tomography of fields, starting with the problem formulation and discussing the essential steps to solve it. The problem yields non-unique solutions that depend on the steps taken to solve it. We also provide simple examples that illustrate the described theory and an extension to geometric-based differential equations of \cite{KyciaSilhan}.

\section{Preliminaries}
\label{Section_Preliminaries}
For readers' convenience, we summarize results related to geometric decomposition \cite{EdelenExteriorCalculus, EdelenIsovectorMethods, ExteriorSystems, KyciaPoincare, KyciaPoincareCohomotopy, CopoincareHamiltonianSystem} and its applications to solving PDEs on manifolds in a local setup \cite{KyciaSilhan, KyciaPoincare, KyciaPoincareCohomotopy, CopoincareHamiltonianSystem}. See also \cite{ExteriorSystems} for the general theory of exterior differential systems.

The geometric decomposition is based on a star-shaped open subset of Euclidean space. This subset, when it fits into a chart, is mapped to a manifold $M$. Therefore, without sacrificing generality, we will focus on an open star-shaped $U\subset \mathbb{R}^{n}$, where $0< n <\infty$ is the dimension of an underlying manifold $M$.

We can define the module $\Lambda^{*}(U)=\Lambda^{0}(U)\oplus\Lambda^{1}(U)\oplus\ldots\oplus\Lambda^{n}(U)$ of exterior differential forms over $\Lambda^{0}(U) = C^{\infty}(U)$. Then the exterior differential $d:\Lambda^{k}(U)\rightarrow \Lambda^{k+1}(U)$ for $0<k<n-1$.

Star-shaped structure with respect to the homotopy center $x_{0}\in U$ allows one to define a homotopy
\begin{eqnarray}
    F:[0,1]\times U\rightarrow U,\\
    F(t,x)=x_{0} + t(x-x_{0}),
\end{eqnarray}
that connects $x\in U$ with the center. The homotopy defines a a curve $t \rightarrow F(t,x)$ and the tangent vector to this curve $\mathcal{K}=(x-x_{0})^{i}\partial_{i}$.

We can define a (linear) homotopy operator as
\begin{eqnarray}
H:\Lambda^{k}(U)\rightarrow \Lambda^{k-1}(U),\\
H\omega = \int_{0}^{1} i_{\mathcal{K}}\omega|_{F(t,x)}t^{k-1}dt,    
\label{eq.HomotopyOperator}
\end{eqnarray}
where $i:\Lambda^{k}(U)\rightarrow \Lambda^{k+1}(U)$ is the insertion operator $i_{\mathcal{K}}\omega(\dots) = \omega(\mathcal{K},\ldots)$.

The linear homotopy operator has many useful properties like nilpotency $H^{2}=0$, and interacts with exterior derivative as $dHd=d$, $HdH=H$. However, the most useful formula is the homotopy invariance formula
\begin{equation}
 dH + Hd = I - s_{x_{0}}^{*},
 \label{Eq_homotopyInvarianceFormula}
\end{equation}
where $s_{x_{0}}^{*}$ is the pullback along the constant map $s_{x_{0}}:x_{0} \hookrightarrow U$. The map $s_{x_{0}}^{*}$ is non-vanishing only on $\Lambda^{0}(U)$, so for $\Lambda^{k}(U)$ for $k>0$ we have from (\ref{Eq_homotopyInvarianceFormula}) and the properties of $H$ and $d$ that
\begin{eqnarray}
 (dH)^{2}=dH, ~ (Hd)^{2} = Hd, \\
 dH + Hd = I.
\end{eqnarray}
This shows that $dH$ and $Hd$ are projection operators. We define
\begin{itemize}
    \item {\textbf{exact} vector space - $\mathcal{E}^{k}(U)=ker(d) = im(dH)$,}
    \item {\textbf{antiexact} module over $C^{\infty}(U)$ - $\mathcal{A}^{k}(U)=ker(H)=im(Hd)=\{\omega\in \Lambda^{k}(U)| H\omega =0, \omega|_{x_{0}}=0\}$.}
\end{itemize}

Summing up, we have the \textit{geometric decomposition}
\begin{equation}
\Lambda^{k}(U)=\mathcal{E}^{k}(U)\oplus \mathcal{A}^{k}(U), ~ k\in\{0,\ldots,n\}.
\end{equation}
This is a local decomposition since it depends on the choice of $x_{0}$ via the $H$ operator \cite{EdelenExteriorCalculus}.

We can also define the metric tensor on $U$ that is $g:TU\odot TU \rightarrow \mathbb{R}$. It can be used to define the correspondence called musical isomorphisms $\_^{\flat}: TU\rightarrow \Lambda^{1}(U)$, i.e., $v^{\flat}=g(v,\_)$; and $\_^{\#}:\Lambda^{1}(U)\rightarrow TU$ defined by the inverse metric. The metric tensor induces also a notion of the Hodge star $\star: \Lambda^{k}(U)\rightarrow \Lambda^{n-k}(U)$, the codifferential (Hodge's star dual to $d$), $\delta= \star^{-1}d\star \eta$, for involutive isomorphism $\eta\omega = (-1)^{k}\omega$ where $\omega\in \Lambda^{k}(U)$. Then the Laplace-Beltrami operator is defined as $\triangle = d\delta + \delta d$, for details see \cite{Thirring, Nakahara}. The Hodge's star dualization can also be applied to the homotopy operator $$h=\eta \star^{-1}H\star$$, see \cite{KyciaSilhan}. Dualization of homotopy invariance formula \cite{KyciaPoincareCohomotopy, CopoincareHamiltonianSystem} is $h\delta+\delta h = I-S_{x_{0}}$, where boundary projection operator is $S_{x_{0}}=\star^{-1}s^{*}_{x_{0}}\star$. Moreover, the geometric decomposition can be dualized: $\Lambda^{*}(U)=\mathcal{C}(U)\oplus \mathcal{Y}(U)$, where coexact vector space is $\mathcal{C}(U)=ker(\delta)=im(\delta h)$ and anticoexact module over $C^{\infty}(U)$ is $\mathcal{Y}(U)=\{\omega \in \Lambda^{*}(U)| \mathcal{K}^{\flat}\wedge \omega=0\quad \omega|_{x=x_{0}}=0 \}=im(h\delta)$.

The above considerations can be extended component-wise to the vector-valued differential forms $\Lambda^{*}(U,V)=\Lambda^{*}(U)\otimes V$ for a finite-dimensional vector space $V$. The vector-valued differential forms naturally fits into framework of associated vector bundle over $U$ with vector space $V$ being the fiber. These forms are equivariant with respect to the structure group $G$ of principal bundle, as discussed in \cite{KyciaSilhan}, see also \cite{Bleecker, Nakahara, LoringTu}. We will not need the framework of associated vector bundles to enforce maximal generality, however, the methods presented here can be without problems adapted to this framework.

The geometric decomposition allows us to solve the exterior differential equations defined on $U$ using methods similar to the ones known from ODE theory \cite{KyciaSilhan}. The common in applications is the covariant exterior derivative
\begin{eqnarray}
    \dcov: \Lambda^{k}(U,V)\rightarrow \Lambda^{k+1}(U,V),\\
    \dcov = d + A\wedge\_,
\end{eqnarray}
where $A\in \Lambda^{1}(U, End(V))$ is a connection one-form. The curvature (of connection is $F=\dcov \dcov= dA+A\wedge A$, and it is a tensorial operator. We can define (inhomogeneous) parallel transport equation for a form $\phi\in \Lambda^{k}(U,V)$ as
\begin{eqnarray}
    \dcov \phi = J,
    \label{Eq.InhParallelTrEq}
\end{eqnarray}
for fixed $J\in \Lambda^{k+1}(U,V)$. The RHS can be decomposed into $J= J_{e} \oplus J_{a}$, where $J_{e}=dHJ$ and $J_{a}=HdJ$. The structure of the solutions of (\ref{Eq.InhParallelTrEq}) is characterized in \cite{KyciaSilhan}. We cite it here for the reader's convenience. First, we define gauge modes that are elements of $\mathcal{E}(U,V)\cap ker(A\wedge\_)$. 

For convergence considerations, we need norms. We define the simplest supremum norms. For $\omega \in \Lambda^{k}(U,V)$ that is $\omega = \omega^{i}_{I}dx^{I}e_{i}$, where $I = (i_{1}, \ldots, i_{n})$, $i_{k}\in\{0,1\}$, and where $dx^{I}=dx_{1}^{i_{1}}\wedge\ldots\wedge dx_{n}^{i_{n}}$, define the norm
\begin{equation}
 ||\omega||_{\infty} = max_{i,I}sup_{x\in K} |\omega^{I}_{i}(x)|,
 \label{Eq.Norm_Vect_form}
\end{equation}
where supremum is taken over $\bar{U}$.

For the connection form $A = [A^{i}_{j,k}(x)dx^{k}]$ the supremum norm is defined as
\begin{equation}
 ||A||_{\infty} = max_{ijk} sup_{x\in K} |A^{i}_{j,k}(x)|.
 \label{Eq.Norm_End_form}
\end{equation}

Now we can focus on the solution of (\ref{Eq.InhParallelTrEq}) for different types of inhomogeneity $J$ of increasing complexity.
We start from the homogeneous case $J=0$.
\begin{Theorem}(Theorem 1 of \cite{KyciaSilhan})
\label{Th_homogenous_solution}
The unique nontrivial solution to the equation
 \begin{equation}
  \dcov \phi =0, ~ k>0
  \label{Eq_homogenous_equation}
 \end{equation}
 with the condition $dH\phi= c\in \mathcal{E}(U,V)\setminus ker(A\wedge\_)$, $c\neq 0$, is given by
 \begin{equation}
  \phi = \sum_{l=0}^{\infty} (-1)^{l} (H(A\wedge \_))^{l} c,
  \label{Eq.Solution_homogenous_k_gt_0}
 \end{equation}
where $c$ is an arbitrary form, $(H(A\wedge \_))^{0}=Id$, and
\begin{equation}
 (H(A\wedge \_))^{l} = \underbrace{H(A\wedge ( \ldots  (H(A\wedge \_ )\ldots )}_{l},
\end{equation}
is the $l$-fold composition of the operator $H\circ A \wedge \_$.

The series in (\ref{Eq.Solution_homogenous_k_gt_0}) is uniformly convergent to the continuous form ($\Lambda^{*,1}(U,V)$) inside the ball $B(x_{0}, r)$ with center $x_{0}$ and the radius $r$ given by
\begin{equation}
 r \frac{||A||_{\infty}}{k} = 1,
 \label{Eq_convergence_homogenous_solution}
\end{equation}
 where the supremum is taken over the ball. Moreover, $d\phi = A\wedge \phi$ is a continuous form.

 The continuous solution is an element of $\Lambda^{k,0}(U,V)\setminus ker(A\wedge\_)$. Two of the solutions differ by an exact gauge mode obtained from (\ref{Eq.Solution_homogenous_k_gt_0}) when $c\in \mathcal{E}(U,V) \cap ker(A\wedge\_)$, i.e., $\phi=c$.
\end{Theorem}
The following case involves the exact $J$.
\begin{Theorem}(Theorem 2 of \cite{KyciaSilhan})
\label{Th_nonhomogenous_solution_exactRHS}
 The unique solution $\phi \in \Lambda^{k}(U, V)\setminus ker(A\wedge\_)$ of
 \begin{equation}
  \dcov \phi  = J_{e}, ~ dHJ_{e}=J_{e},
  \label{Eq_nonhomogenous_covariant_equation}
 \end{equation}
for $A\in \Lambda^{1}(U, End(V))$, $J_{e}\in \mathcal{E}^{k+1}(U,V)$ for $k>0$, with $dH\phi=c\in \mathcal{E}(U,V)\setminus ker(A\wedge\_)$ is
\begin{equation}
 \phi=\phi_{H} + \phi_{I}, \quad \phi_{I}=\sum_{l=0}^{\infty}(-1)^{l} (H(A\wedge\_))^{l} HJ_{e},
 \label{Eq_solution_nonhomogenous_k_gt_0}
\end{equation}
where $\phi_{H}$ is a solution of homogenous equation ($J_{e}=0$) given in Theorem \ref{Th_homogenous_solution}.

The series in (\ref{Eq_solution_nonhomogenous_k_gt_0}) is convergent to the continuous solution in the ball $B(x_{0},r)$, where the radius is given by $r \frac{||A||_{\infty}}{k}=1,$, where the supremum norm is taken over the closed ball.

When $HJ_{e}\in \mathcal{E}^{k+1}\cap ker(A\wedge\_)$, then the solution reduces to
\begin{equation}
 \phi = \phi_{H}+HJ_{e}.
\end{equation}
\end{Theorem}

Note that the solution (\ref{Eq_solution_nonhomogenous_k_gt_0}) can be written as
\begin{eqnarray}
    \phi = \frac{1}{I+HA\wedge\_}(c + HJ_{e}),
    \label{Eq.ExactInhomogenitySolution}
\end{eqnarray}
where the operator fraction is understood in terms of a formal series expansion. It is convergent when within the radius of convergence.

The fraction can also be understood in terms of Mikusinski's ring \cite{KyciaSilhan}. Denote the denominator as 
\begin{eqnarray}
    G = I+ HA\wedge\_.
\end{eqnarray}
We have 
\begin{Proposition}(Proposition 8 of \cite{KyciaSilhan}
\label{Proposition_no_zero_divisions}
 If $\dcov\phi=0$ then we have
 \begin{equation}
  G\phi =0 \Rightarrow \phi=0.
 \end{equation}
\end{Proposition}
Therefore, $G$ has no zero divisors on the solutions of the homogeneous equation.

The most general case is presented by the following
\begin{Theorem} (Theorem 4 of \cite{KyciaSilhan})
\label{Th_FullInhomogenous_parallelTransportEquation}
 The solution of the inhomogeneous covariant constancy equation
 \begin{equation}
  \dcov \phi = J,\quad \dcov = d + A\wedge \_,
  \label{Eq.FullInhomogenous_CovariancyConstantEquation}
 \end{equation}
where $\phi\in\Lambda^{k}(U, V)$, $A\in \Lambda^{1}(U,End(V))$, $J\in \Lambda^{k+1}(U,V)$, $k>0$ is given by
\begin{equation}
 \phi = \phi_{1}+\phi_{2}+\phi_{3},
\end{equation}
where $\phi_{1}$ fulfils
\begin{equation}
 \dcov \phi_{1}=J_{e} - d(\phi_{2}+\phi_{3}),
\end{equation}
$\phi_{2}$ fulfils
\begin{equation}
 A\wedge \phi_{2} = J_{a},
 \label{Eq_Ja_condition}
\end{equation}
 where $J_{e}:=dHJ$ is the exact part of $J$, and $J_{a}:=HdJ$ is the antiexact part of $J$.
 Finally, we have an arbitrary choice for $\phi_{3}\in \ker(A\wedge\_)$. Moreover, $A\wedge \phi_{1}\in \mathcal{E}^{k+1}(U, V)$ and $A\wedge\phi_{2}\in \mathcal{A}^{k+1}(U,V)$.

 If the equations cannot be solved, then there is no solution of (\ref{Eq.FullInhomogenous_CovariancyConstantEquation}). The only equation that imposes constraint is (\ref{Eq_Ja_condition}) and can be fulfilled only when $J_{a}\in Im(A\wedge\_)$.
\end{Theorem}

For now, we consider forms with coefficients that are smooth functions, as in classical Differential Geometry. However, we will also need Schwartz's theory of distributions in terms of exterior differential forms. G. de Rham developed such forms \cite{deRham} under the name of currents. We need the version for vector-valued exterior differential forms. Assume that $j:B \hookrightarrow U$ is a closed (smooth) boundary with interior $int(B)$. Then the distributional exterior derivative of a form $\Phi$ with distributional (in terms of Schwartz) coefficients is defined by means of Stokes' theorem
\begin{equation}
    \int_{int(B)} \phi\wedge d\Phi = \int_{B} \phi\wedge\Phi - \int_{int(B)}d\phi\wedge \Phi,
\end{equation}
where $\phi \in \Lambda^{n-k}_{0}(\bar{B}, V^{*})$ - forms vanishing on $B$ with the values in dual of $V$ with natural pairing with $\Phi\in\Lambda^{k}(\bar{B}, V)$. This is evident from the definition $\Lambda^{k}(\bar{B}, V)=\Lambda^{k}(U)\otimes V$ and the definition of duals to the components of the tensor product.

\section{Results}
We consider an open star-shaped subset $U$ of $n$-dimensional Euclidean space $\mathbb{R}^{n}$. We select a center $x_{0}\in U$ and then the homotopy $F(x,t)=x_{0} + t(x-x_{0})$ for $t\in[0,1]$ and $x \in U$. We also consider a boundary $B \in U$ that is a regular convex closed surface if $dim(U)>1$ and a $2$-point boundary condition for $dim(U)=1$. This means that every point $x\in B$ can be connected with the center $x_0$.

Let moreover $V$ be a vector space and define $V$-valued exterior differential module $\Lambda^{*}(U,V)$ over $U$. For injection $j:B \hookrightarrow U$, we can introduce the notation
\begin{equation}
    \phi|_{B} := j^{*}\phi, \quad \phi \in \Lambda^{*}(U,V).
\end{equation}

\begin{Definition}
We consider a covariant tomography/IBVP that is to find a solution $\phi \in \lambda^{*}(U, V)$ of
\begin{eqnarray}
 \dcov \phi = J, \label{Eq.A_CovariantTransport} \\
  \phi|_{B} = \alpha \label{Eq.B_BoundaryValue}
\end{eqnarray}
for $J \in \Lambda^{*}(U,V)$ and $\alpha \in \Lambda^{*}(B,V)$ and $A\in\Lambda^{1}(U,L(V,V))$.

We can ask the following question: Knowing $\alpha$, can we recover:
\begin{itemize}
    \item {$J$, knowing $A$? - current tomography;}
    \item {$A$, knowing $J$? - gauge field tomography;}
    \item {$J$ and $A$? - current and gauge fields tomography;}
\end{itemize}

The geometry of the problem is presented in Fig. \ref{fig:Geometry}.
\begin{figure}
    \centering
    \includegraphics[width=0.5\linewidth]{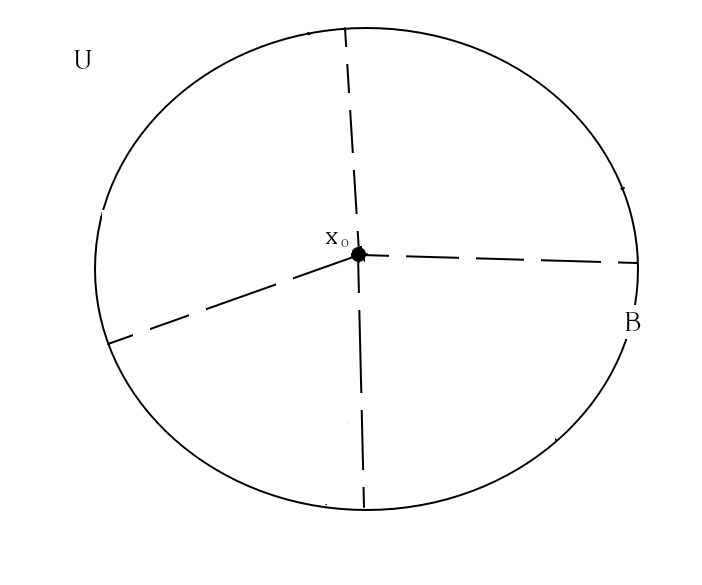}
    \caption{Geometry of the problem.}
    \label{fig:Geometry}
\end{figure}
\end{Definition}

The method of solving the covariant tomography problem requires matching and connecting two 'loose ends':
\begin{enumerate}
    \item {\textbf{Extension problem}: Extend boundary condition $\alpha$ from (\ref{Eq.B_BoundaryValue}) to the whole interior of $B$, calling it $\Phi$. This process can be done in multiple ways, discussed below, and resembles the pseudoinverse of the 'trace' operator given by the restriction of the field $\phi$ to the boundary. The extension, however, does not have to fulfill (\ref{Eq.A_CovariantTransport}).}
    \item {\textbf{Projection problem}: Project $\Phi$ to the solution of (\ref{Eq.A_CovariantTransport}).}
\end{enumerate}
We will consider the solution of these two partial problems in the following subsections.

\subsection{Extension problem}
We now discuss the ways of extending $\alpha$ to the whole interior of $B$. This yields the map $\tilde{\Phi}:\Lambda^{*}(B,V)\rightarrow \Lambda^{*}(U,V)$,
\begin{equation}
    \tilde{\Phi} (\alpha) = \Phi.
\end{equation}
It can be constructed in various ways. We discuss three of them:
\begin{enumerate}
    \item {Constant extension along the rays - radial extension.}
    \item {Heat equation extension.}
    \item {Harmonic extension.}
\end{enumerate}

For extension by the constant value at the boundary along the ray - \textit{radial extension} - we have to use the homotopy $x = F(x_{b},t_{x})=x_{0} + t_{x}(x_{b}-x_{0})$ that connects $x_{b}\in B$ with the center $x_{0}$ and such that $x$ is on the ray $[x_{0}, x_{b}]$ for the homotopy parameter $t_{x}\in [0,1]$. We assume that $x$ is on the ray.
\begin{equation}
    \tilde{\Phi}(\alpha)(x) = \alpha(x).
\end{equation}
The fundamental problem of this approach is the possible discontinuity at $x_{0}$. That behavior can be detected by evaluating the current
\begin{equation}
  J_{ext} = \dcov \Phi.
  \label{Eq.ExtensionCurrent}
\end{equation}
The exterior derivative in $\dcov$ in this case must be understood as a distributional derivative. In the case of discontinuity of $\Phi$, the current $J_{ext}$ is a singular distribution that compensates it. It also measures the non-covariant constancy of $\Phi$. We obtain
\begin{Proposition}
    The radial extension maps $C^{k}(B)$ boundary data onto $C^{0}(U\setminus \{x_{0}\})$ interior solution.
\end{Proposition}

The \textit{heat extension} approach requires to introduce a metric on $U$. We can select the Euclidean metric $g$ for defining the Hodge's star $\star$, codifferential $\delta$, and the Laplace-Beltrami operator $\Delta = d\delta + \delta d$. That allows us to solve the BVP problem
\begin{equation}
\left\{
\begin{array}{c}
     \frac{\partial}{\partial t} \Phi_{h}(x,t) = \Delta \Phi_{h}(x,t),  \\
     j^{*}\Phi_{h} = \alpha.
\end{array}
\right.
\end{equation}
The initial data for $int(B)$ can be set arbitrarily, e.g., as $\Phi_{h}(x,0)=0$ for $x \in int(B)$. For practical purposes, we must select the time of evolution as finite $t=T$. The resulting one-parameter family of extension operators is
\begin{equation}
\tilde{\Phi}_{T}(\alpha)(x) = \Phi_{h}(x,T).    
\end{equation}
The diffusion problem generally smooths the initial conditions instantaneously \cite{HeatRobinson}. The current (\ref{Eq.ExtensionCurrent}) is now smooth, and its nonzero values indicate compensation currents of extension. Considering the heat equation property we obtain
\begin{Proposition}
    The heat extension maps $L^{2}(B)$ boundary value onto $C^{\infty}(U \times (0,T])$ interior solution.
\end{Proposition}

The last case - \textit{harmonic extension} - is formally the previous case with $t\rightarrow\infty$. More precisely, as before, we define a metric structure that results in the Laplace-Beltrami operator. Then the extension problem is equivalent to solving 
\begin{equation}
\left\{
\begin{array}{c}
    \triangle \Phi =0 \\
    j^{*}\Phi = \alpha.
\end{array}
\right.
\end{equation}
Knowing the solution $\Phi$, the extension operator $\tilde{\Phi}(\alpha) = \Phi$.
As harmonic solutions are smooth when starting from smooth boundary data, the extension current (\ref{Eq.ExtensionCurrent}) will be regular. We have
\begin{Proposition}
    The harmonic extension maps $C^{\infty}(B)$ boundary value onto $C^{\infty}(U)$ interior solution.
\end{Proposition}

Summing up, the non-smooth radial extension yields an extension current that is, in general, distributional. It is straightforward, however, decrease smoothens. The other approaches require solving a PDE that produces a smooth solution, and as a result, the current is smooth.

In the following subsection, we present the restoration of the parallel transport equation using the extension $\Phi$.

\subsection{Projection problem}
Having $\Phi$ in $int(B)$, we now check under what conditions it is the solution $\phi$ of $\dcov \phi = J$ that depends on $\Phi$. Moreover, we have internal properties that are characterized by $A$ and $J$.

\noindent\textbf{Case - $A$ fixed:} We have the following: 
\begin{Theorem}
    Let $U$ be a star-shaped subset and $A$ a fixed connection one-form. Given boundary data $\alpha$, a unique interior current $J$ can be recovered (up to gauge mode) trough the mapping $J=\dcov \Phi$, where $\Phi$ is the harmonic extension of $\alpha$.
\end{Theorem}
\begin{Proof}
The connection form $A$ is fixed, therefore $\dcov$ is also fixed. We compute the current $J_{ext} = \dcov \Phi$. We consider three cases.
\begin{itemize}
    \item {If $J_{ext}=0$ the field $\Phi$ is already covariant constant. We can restore its form by setting $c = G\Phi$, and the solution is, according to Theorem \ref{Th_homogenous_solution}, given by $\Phi = \frac{1}{G}c$.}
    \item {If $J_{ext} = J_{e}$ is exact, $J_{ext}=dHJ_{ext}$, then using (\ref{Eq.ExactInhomogenitySolution}), we can compute $c = G\Phi - HJ_{e}$, and represent the solution as $\Phi = \frac{1}{G}(c+HJ_{e})$.}
    \item {In the remaining case, $J_{ext} = J_{e}\oplus J_{a}$ with $J_{a}\neq 0$. Using Theorem \ref{Th_FullInhomogenous_parallelTransportEquation}, we then have the solution in the form $\Phi = \phi_{1}+\phi_{2}+\phi_{3}$, where three components are determined by
\begin{eqnarray}
    A\wedge \phi_{1}=J_{a}, \\
    \phi_{2} \in ker(A\wedge\_),\\
    \dcov \phi_{3}=J_{e}-d(\phi_{1}+\phi_{2}) =: \tilde{J_{e}}.
\end{eqnarray}
The first equation is an algebraic one, the second is an arbitrary field that can be used for adjustments, and $\phi_{3}$ can be computed as in the previous case, we compute $c = G(\Phi-\phi_{1}-\phi_{2}) - H\tilde{J_{e}}$. Then $\phi_{1}=\frac{1}{G}(c+H\tilde{J_{e}})$.}
\end{itemize}
\end{Proof}

\noindent\textbf{Case - $J$ fixed:} We have:
\begin{Theorem}
    Given an interior current $J$ and an extended field $\Phi$ form the boundary field $\alpha$, the connection $A$ is (non-uniquely) determined by the algebraic relation $A\wedge\Phi=J-d\Phi$.
\end{Theorem}
\begin{Proof}
When $J$ is fixed we have to solve the problem for $A$ that yields algebraic equations of the form
\begin{eqnarray}
    A\wedge \Phi = J-d\Phi.
\end{eqnarray}
Using geometric decomposition with $J_{e}=dHJ$, $J_{a}=HdJ$, $(A\wedge\Phi)_{e}=dH(A\wedge\Phi)$, $(A\wedge\Phi)_{a}=Hd(A\wedge\Phi)$, we get
\begin{eqnarray}
    (A\wedge \Phi)_{e}=J_{e}-d\Phi, \\
    (A\wedge\Phi)_{a} = J_{a}.
\end{eqnarray}
We also note that $A$ is determined up to $ker(\Phi\wedge\_)$.    
\end{Proof}

\noindent\textbf{Case - $A$, $J$ to be determined:} When both $A$ and $J$ are to be determined, then some additional conditions should be imposed on $J$, $A$, $\phi$, or the curvature $F$. The general idea is to construct a functional that penalizes values outside the optimal selection \cite{InverseProblemsInDE}. In the parallel transport equation, the solution formula is expressed by the expansion of $G^{-1}$ and has the radius convergence depending on $||A||$. Moreover, the great change in $A$ introduces (via $\dcov$) a large change in $J$, and therefore $||F||$ is also large in this case. One of the possible functional adjuncts to the problem defined by (\ref{Eq.A_CovariantTransport}) and (\ref{Eq.B_BoundaryValue}) is 
\begin{eqnarray}
    min_{A} a||A||^{2}+b||F||^{2}, ~ a,b \in \mathbb{R}_{+},
\end{eqnarray}
where $A$ is taken from some suitable closed subset of $\Lambda^{1}(U, End(V,V))$. The first term can be understood as Tikhonov regularized \cite{InverseProblemsInDE}, and the second term is related to the Yang-Mills term.

\subsection{Convergence, stability, and non-uniqueness}
In the proposed scheme for covariant tomography, there are two critical issues that we discuss in this subsection.

\textbf{Convergence}: The method for constructing local solutions of covariant transport has underlying the circle of convergence of the radius $\frac{k}{||A||}$ for $k$ being the grade of $\phi$, as described in theorems cited in Section \ref{Section_Preliminaries}. If the connection norm is large, then the radius of convergence can be smaller than the whole $int(B)$. Then, the method of splitting the domain into smaller ones and then matching the partial domains must be applied. The matching can be done extending the $U$ to associated vector bundle with vector space $V$, and principal bundle with fiber being group $G$, see \cite{Nakahara, Bleecker}. Then by subdividing $U$ into star-shaped open family of overlapping sets $\{U_{i}\}_{i}$ with homotopy centers $x_{0,i}\in U_{i}$. If $U_{i}\cap B\neq \emptyset$ then we solve in this domain IBVP with the parts of $\alpha$ that are defined on the boundary. The sets $U_{i}$ that have no common parts with $B$, inherit the boundary values from the sets $U_{j}$ that $U_{i}\cap U_{j} = \emptyset$, and these values are used to solve IBVP. This can be done by iterative numerical procedure for complicated domains. For matching gauge field $A$ the gauge transformation can be used in the form $A_{j} = gA_{i}g^{-1}+g(dg)^{-1}$, where $g$ is an element of a Lie group associated with the gauge field \cite{Nakahara} that transforms the field $A_{i}$ from $U_{i}$ to $A_{j}$ on $U_{j}$. At the same time the field should transform as $\phi_{j}= \phi_{i}g$. When $U$ has complicated topological structure, the obstructions in matching can arise leading to non-trivial cohomology group \cite{LoringTu}.

\textbf{Stability of extension operator}: The extension operator $\tilde{\Phi}$ yields the $J_{ext}$ with different regularity. For radial extension, the possible discontinuity of the extended field at $x_{0}$ may arise in the singularity of $J_{ext}$. In the framework of Schwartz's theory of distributions, this discontinuity manifests as a singular distribution in $J_{ext}$. Moreover, the small perturbations of the boundary value $\alpha$ propagate to $int(B)$. On the other hand, extension using the heat equation and harmonic forms has smoothing properties for boundary values, but comes at an additional cost of selecting the metric tensor. 

\textbf{Non-uniqueness}: The lack of a unique 'pseudoinverse' for the trace operator $j^{*}$ (restriction to the boundary) allows for multiple interior extensions $\Phi$ of the same boundary value $\alpha$. The possible non-uniqueness arises also from the existence of the gauge modes $\mathcal{E} \cap ker(A\wedge\_)$ and the fields from $ker(A\wedge\_)$, which results from the solution method based on geometric decomposition \cite{KyciaSilhan}.

\subsection{Tomography for geometric-based equations}
When $U$ carries metric tensor, the codifferential $\delta$ can be constructed as well as the (Hodge's star) dual to covariant derivative for gauge field $A_{i}$, which we denote $D_{A_{i}}:=d+A_{i}\wedge\_:\Lambda^{k}(U,V)\rightarrow\Lambda^{k+1}(U,V)$ that is
\begin{eqnarray}
    \reflectbox{D}_{X_{i}}:=\delta+i_{X_{i}}:\Lambda^{k+1}(U,V)\rightarrow \Lambda^{k}(U,V),
\end{eqnarray}
see \cite{KyciaSilhan} for details. The theory of solving the inhomogeneous contravariant constancy equation
\begin{eqnarray}
    \reflectbox{D}_{X_{i}} \phi = J
\end{eqnarray}
is described in \cite{KyciaSilhan} and results from the theorems of this paper cited in Section \ref{Section_Preliminaries}. They have the same form with replacing 
\begin{eqnarray}
    d & \leftrightarrow & \delta, \\
    A\wedge\_ & \leftrightarrow & i_{X},
\end{eqnarray}
which results from Hodge's star(/metric) dualization \cite{deRham}.

These two operators can be composed in multiple ways, yielding so-called geometric-based differential equations. As discussed in \cite{KyciaSilhan}, the equation containing a composed operator can be decomposed into covariant and contravariant parallel transport equations exactly in the same way as higher-order ODEs can be rewritten as a system of first-order ODEs. This is provided by the following:
\begin{Algorithm}[Tower Method]
\label{Alg_Gometric-BasedDiffEq}
    In order to solve the geometric-based equations of the form 
    \begin{equation}
    \left\{
        \begin{array}{c}
        D \phi = J, \\
        j^{*}\phi=\alpha
        \end{array}
    \right.
        \label{Eq.gemetricaly-basedDiffEq}
    \end{equation}
    with a fixed boundary data $j^{*}\phi=\alpha$, where $D$ is the $k$ composition of $D_{A}$ and $\reflectbox{D}_{X}$ operators 
    \begin{enumerate}
        \item {\textbf{Decompose} the equation into a system of coupled first-order equations
        \begin{equation}
        \left\{
        \begin{array}{c}
            D_{k}\phi_{k} = J \\ 
            D_{k-1}\phi_{k-1}=\phi_{k} \\
            \ldots \\
            D_{2} \phi_{2} = \phi_{3} \\
            D_{1} \phi_{1} = \phi_{2} \\
            j^{*}\phi_{k}=\alpha, 
        \end{array}
        \right.
        \label{Eq.TowerGeometric-BasedDiffEq}
        \end{equation}
        where $D_{i}$ is either $D_{A}$ or $\reflectbox{D}_{X}$ operator.
        }
        \item {\textbf{Solve} the IBVP $D_{k}\phi_{k}=J$, $j^{*}\phi_{k}=\alpha$ as described in the previous section.}
        \item {\textbf{Propagate} the solution $\phi_{k}$ for other $\phi_{i}$ solving recursively remaining parallel/covariant transport equations.}
    \end{enumerate}
\end{Algorithm}
We therefore have
\begin{Theorem}
    For the IBVP problem (\ref{Eq.gemetricaly-basedDiffEq}) to be solved if and only if associated tower of first-order equations (\ref{Eq.TowerGeometric-BasedDiffEq}) is sequentially solvable.
\label{Th_TowerTheorem}
\end{Theorem}

We illustrate the algorithm \ref{Alg_Gometric-BasedDiffEq} by simple example
\begin{Example}
Consider
\begin{eqnarray}
    D\reflectbox{D} \phi=J, \\
    j^{*}\phi=\alpha
    \label{Eq.GeometricBasedDEExample}
\end{eqnarray}
which can be decomposed as 
\begin{eqnarray}
    \reflectbox{D} \phi_{1} = \phi_{2},\\
    D\phi_{2}=J, \\
    j^{*}\phi_{2}=\alpha.
\end{eqnarray}
The last two equations form an IBVP for its own, and should be solved in the first place using the above methods. Then the first equation forms the IBVP with fixed current $\phi_{2}$. 
\end{Example}

In general, we get a tower of coupled equations, and the one for the field in the original equation will be connected to the boundary values. The solution method propagates the solution of the first equation in this tower sequentially to the last one.

\subsection{Examples}

We provide two simple examples of IBVP with $U=\mathbb{R}$, $B=\{0, 1\}$, so with $int(B)=(0,1)$. The center of homotopy is $x_{0}=\frac{1}{2}$. Since we will use harmonic extension, we introduce a simple Euclidean metric with $g=[1]$.

\begin{Example}[$\Lambda^{0}$ on $(0,1)$]
\label{Ex.1}
Consider boundary values $\phi(0)=1$ and $\phi(1)=2$. Using harmonic extension, we solve the Laplace equation $\frac{d^{2}\Phi}{dx^{2}}=0$. We get $\Phi(x)=x+1$.

Assume that $A=0$, so $G=I$. Then the current is $J_{ext}=d\Phi = dx$ that is exact. Therefore, $HJ = x-x_{0}$ and $c = 1+x_{0}$.

Now Assume that $A=f(x)dx$. Then the current is $J_{ext}= (1+f(x)(1+x))dx$.

When $J=g(x)dx$ is fixed and $A=f(x)dx$ arbitrary, then $f(x)=\frac{g(x)-1}{x+1}$. For example, if we fix $J=0$ then it enforces the shape of the connection to $A = \frac{-1}{x+1}$.

In general, when $A=f(x)dx$ and $J=g(x)dx$ are arbitrary, then it yields only a constraint $1+(1+x)f(x)=g(x)$. An additional condition must be introduced to fix the explicit form of $J$ and $A$, e.g., an additional functional that is under optimization involving $A$ and $J$.
\end{Example}
\begin{Example}[$\Lambda^{1}$ on $(0,1)$]
Consider now boundary conditions given by $1$-forms: $\phi(0)=dx$, $\phi(1)=2dx$. Solving the harmonic extension problem yields $\Phi = (x+1)dx$.

The field $\Phi$ is automatically covariant constant with $J=0$ and arbitrary $A$ due to the low dimensionality of space.
\end{Example}

Extending this example to higher dimensions, we have
\begin{Corollary}
For $dim(U)=n$ the top form is automatically covariant constant with $J=0$ and arbitrary $A$.
\end{Corollary}

We now provide example of central singularity during the radial extension.
\begin{Example}[$\Lambda^{0}$ on $(0,1)$ - radial extension]
Consider again boundary values as in Example \ref{Ex.1}. The radial extension yields 
\begin{eqnarray}
    \Phi(x) = 1 + \theta\left(x-\frac{1}{2}\right), x\in [0,1],
\end{eqnarray}
where $\theta(x)$ is the Heaviside unit jump from $0$ to $1$ at $x_{0}=0$. It is not important what is the value at $x_{0}$ since it is a zero measure set.

Taking $A=f(x)dx$, we can compute $J_{ext}= \delta\left(1-\frac{1}{2}\right)dx + f(x)(H\left(x-\frac{1}{2}\right)+1)dx$, where $\delta(x)$ is the Dirac's delta distribution that generates singularity at the homotopy center. We can even extract the terms related to the singular distribution and a simple discontinuity.
\end{Example}

In the next examples we consider IBVP for Maxwell equations \cite{Nakahara, Thirring}. 
\begin{Example}[Electromagnetic potential in $\mathbb{R}^{3}$ knowing $J$.]
Consider a star-shaped $U\subset \mathbb{R}^{3}$ and a spherical boundary $B$ centered at $x_{0}$ that is the center of homotopy for $U$.

The Maxwell equation on $U$ can be written
\begin{eqnarray}
    dA =F, \\
    \delta F = J, 
    \label{Eq.Maxwell}
\end{eqnarray}
that is a form of a coupled second-order geometric-based differential equation. 

We will construct the procedure for solving for $A$ from the knowledge of a field $F$ at $B$, i.e., we fix $\alpha \in \Lambda^{1}(B,V)$ and a fixed form of $J$.

First, we have to note that the current $J$ is coexact ($\delta J=0$ or $J = \delta  hJ$), which expresses the conservation of current. Therefore, we get $F = c_{1} + hJ$ for an arbitrary coexact $1$-form $c_1$.

Then the projection requires $F$ to be exact ($dF = 0$). This place the constraints on $c_{1}$, that is, $d(c_{1}+hJ)=0$. When fulfilled, the first Maxwell equation yields $ A = c_{2} + H(c_{1}+hJ)$, where $c_{2}$ is an arbitrary exact form.

We observe that the solution for $A$ is non-unique, as it depends on the arbitrary $c_{2}$. Moreover, the extension procedure is more straightforward since we can use the geometric decomposition for the $\delta$ operator directly as $\dcov = d$ in this case.
\end{Example}

\begin{Example}[Potential in $\mathbb{R}^{3}$ knowing $J$ and electromagnetic field at boundary.]
Using the setup as in the previous example, fix $J$ and the boundary value $j^{*}F=\alpha$.

Then, as before, the general solution of (\ref{Eq.Maxwell}) in the form $F = c_{1} + hJ$ for arbitrary coexact $c_{1}$. We have to impose two constraints on $F$, that yields in the choice of $c_{1}$:
\begin{eqnarray}
    j^{*}F = \alpha, \\
    dF = 0.
\end{eqnarray}
Providing the $c_{1}$ can be selected in such a way that it yields, as before, $ A = c_{2} + H(c_{1}+hJ)$, where $c_{2}$ is an arbitrary exact form.
\end{Example}

\section{Conclusions}
This research establishes a robust framework for solving Inverse Boundary Value Problems (IBVPs) through a methodology termed 'covariant tomography'. By integrating geometric decomposition with the linear homotopy operator, we have successfully bridged the gap between boundary restrictions and interior field configurations. This approach provides a systematic way to recover interior currents and gauge potentials within star-shaped domains.

A significant novelty of this work is the introduction of Algorithm \ref{Alg_Gometric-BasedDiffEq}, which decomposes higher-order, geometric-based differential equations into a 'tower' of coupled first-order systems. This allows for the tomographic reconstruction of fields, such as solutions of the Maxwell equations, through sequential parallel transport equations. This framework is formally supported by Theorem \ref{Th_TowerTheorem}, which establishes a solvability criterion: a higher-order IBVP is solvable if and only if its associated first-order tower is sequentially solvable.

While the method offers an algorithmically straightforward local solution via formal series expansion, it is subject to specific constraints:
\begin{itemize}
    \item {\textbf{Convergence}: The solution is limited by a finite radius of convergence dependent on the connection norm.}
    \item {\textbf{Non-uniqueness}: The presence of gauge modes inherent in the geometric decomposition can lead to non-unique interior configurations.}
    \item {\textbf{Regularity}: The smoothness of recovered currents is directly influenced by the choice of extension technique-radial, heat equation, or harmonic.}
\end{itemize}

By grounding these results in the historical progress in multidimensional inverse problems - referencing the foundational work of Gelfand \cite{Gelfand} and Novikov \cite{Novikov} - this research opens new path for applying exterior calculus to field reconstruction in physics and engineering.

\section*{Acknowledgments}
The author declares no conflict of interest.\\
No data is attached to the presented research.\\
The work was partially funded by the EU Horizon Europe MSCA grant No 101183077. \\
The author acknowledges the support of the COST CA24122 and CA22123 actions.




\end{document}